\documentclass[a4paper,11pt]{article}
\usepackage[a4paper,margin=1in]{geometry}
\usepackage[T1]{fontenc}
\usepackage{lmodern}
\usepackage{setspace}
\usepackage{amsmath,amssymb,amsfonts}
\usepackage{braket}
\usepackage{graphicx}
\usepackage{tikz}
\usepackage{quantikz}
\usepackage{subcaption} 
\usepackage{booktabs}
\usepackage{makecell}
\usepackage{placeins}
\usepackage{xcolor}
\usepackage[hidelinks]{hyperref}
\usepackage{authblk}
\usepackage{cite}
\title{Curvature-Aware Edge Dynamic GNNs for Low-$p_T$ Track Reconstruction at HL-LHC}
\author[1]{Syed Haider Ali}
\author[2,3]{Nouhaila Innan}
\author[2,3]{Muhammad Shafique}

\affil[1]{Institute of Physics, Faculty of Science and Technology, University of Debrecen, Egyetem tér 1, H-4032 Debrecen, Hungary}
\affil[2]{eBrain Lab, Division of Engineering, New York University Abu Dhabi, PO Box 129188, Abu Dhabi, UAE}
\affil[3]{Center for Quantum and Topological Systems, NYUAD Research Institute, New York University Abu Dhabi, UAE}

\date{}

\begin{document}

\maketitle
\begin{abstract}
Particle track reconstruction in high-energy collider experiments is a computationally demanding combinatorial problem, traditionally solved by Kalman filters that scale poorly with increasing detector occupancy. Graph Neural Networks (GNNs) have recently emerged as a promising alternative. In this work, we introduce the \textbf{Manifold Helical-IN}, a novel Interaction Network architecture that explicitly embeds detector spacepoints onto a helical manifold chart using curvature-aware features. This geometric projection mathematically unrolls the tightly curling trajectories of low-momentum particles. Consequently, the network can distinguish true physical tracks from combinatorial noise with good precision. Evaluated on the TrackML challenge dataset, the proposed Manifold Helical-IN achieves a reconstruction efficiency of \textbf{0.9675} in the experimentally critical $p_T > 0.5$ GeV regime, approaching recently reported Transformer-based approaches. Crucially, the manifold embedding yields a fake track rate of \textbf{0.0037}. This represents a 13-fold improvement over the baseline Helical-IN (0.0508). Furthermore, the model maintains good track coherence, as quantified by our proposed Mean Hit Recovery Fraction (MHRF) of 0.858, and achieves a highly efficient inference latency of approximately \textbf{150 ms per event} on a single NVIDIA A100 GPU. These results represent a critical step toward real-time soft particle tracking for the High-Level Trigger (HLT). By operating within practical HLT latency budgets, this geometry-aware approach provides a pathway to recover soft, anomalous Beyond Standard Model signatures such as unclustered energy patterns (SUEPs), that routinely evade standard hardware filters.

\end{abstract}

\section{Introduction}

Charged particle track reconstruction is fundamental to data analysis in high-energy physics. At the Large Hadron Collider, collisions generate thousands of particles that leave hits in layered detectors. Traditional methods, such as combinatorial Kalman filters, perform well in low-pileup conditions but become computationally expensive as event complexity grows \cite{strandlie2010pattern}. The High-Luminosity LHC, expected to begin operations in mid 2030s, will increase luminosity by a factor of five to ten. This will raise average pileup to 140--200 interactions per bunch crossing\cite{cmsHLLHC,CMS2020Phase2}, producing hit densities that overwhelm classical algorithms.

Graph neural networks offer a promising alternative. They represent hits as nodes and possible connections as edges, allowing learned pattern recognition in irregular geometries . The Exa.TrkX\cite{Ju2021ExaTrkX,ju2021graph} project has shown that interaction networks can scale to large datasets. Recent work, such as EggNet\cite{Chan2025EggNet}, has reported efficiencies around 0.96 under TrackML benchmark conditions \cite{Kiehn2019TrackML} dataset through iterative graph refinement. However, its multi-step evolution with repeated kNN reconstruction and attention passes makes it computationally intensive. More importantly they have benchmarked and investigated their model performance mostly on tracks from $P_T > 900 MeV$. More research is required in the tracks in the lower $P_T$ domain in order to make a functional GNN based tracker for the HL-LHC.

At the HL-LHC, hardware-based Level-1 (L1) triggers will impose strict transverse momentum thresholds, typically $p_T > 2$ GeV, to satisfy severe bandwidth constraints \cite{Holmberg:2844896}. This rigid cutoff effectively renders soft, diffuse Beyond Standard Model (BSM) signatures such as Soft Unclustered Energy Patterns (SUEPs)\cite{CMS2024SUEP} invisible to standard track-triggers \cite{DiPetrillo2023}. However, L1 systems can flag these anomalous events using global observables like the scalar sum of transverse energy ($H_T$) and pass them to the software-based High-Level Trigger (HLT) \cite{DiPetrillo2023}. Once accepted by the L1 trigger, events are processed by the software-based HLT, which must operate within stringent real-time computing and throughput constraints at the HL-LHC\cite{CMSPhase2HLT} . Standard CPU-based tracking algorithms struggle under the massive combinatorics of these dense, tightly curling soft tracks, often discarding them as noise\cite{lst_chep2023}. Consequently, there is a critical need for highly parallelized, real-time algorithms capable of reconstructing tracks down to 500 MeV at the HLT to prevent these exotic signatures from being permanently deleted.

We address this computational bottleneck through explicit geometry-aware encoding. Rather than asking the network to learn curvature from raw coordinates, we embed hits in a helix-aligned local chart and promote helical differentials to first-class features. Our approach leverages the physical constraint that charged particles follow helical paths in solenoidal magnetic fields\cite{pdg2024}, encoding this prior directly into the graph representation through manifold-aligned coordinates and geodesic-aware edge weighting. In the present study, truth-level $p_T$ is utilized strictly as an idealized structural proxy as a proof of concept. It is set up to be replaced by dynamic curvature estimation calculated directly during graph message passing that will ultimately guide the adaptive search radius without relying on external hardware momentum stubs. We present a controlled architectural study of this physics-informed GNN against baseline variants. While neither model is positioned as a standalone tracker, we analyze their speed-coherence trade-offs to demonstrate how geometric priors can enable robust, high-throughput reconstruction in the combinatorially challenging low-$p_T$ domain.

The paper proceeds as follows. Section~\ref{sec:prelim} details the dataset, evaluation protocol, and graph formulation. Section~\ref{sec:methodology} describes the manifold encoding and interaction blocks. Section~\ref{sec:results} reports controlled benchmarks, momentum-dependent behavior, and regional performance splits. Section~\ref{sec:conclusion} discusses hybrid integration pathways and next steps toward noise-robust deployment.

\section{Preliminaries}
\label{sec:prelim}

\subsection{The TrackML Dataset}
The TrackML challenge, introduced in 2018, provides a realistic simulation of particle tracking in a generic LHC-like detector. The dataset is designed to mimic the conditions expected at the High-Luminosity LHC (HL-LHC), featuring high hit densities and complex event topologies \cite{Kiehn2019TrackML}. Events are generated from proton-proton collisions at a center-of-mass energy of 13 TeV, using a fast simulation that approximates detector responses without full Geant4\cite{Geant4,Kiehn2019TrackML} modeling to enable large-scale production. Each event includes a hard-scattering process overlaid with an average of 200 minimum-bias interactions (pileup), following a Poisson distribution, resulting in approximately 10,000 reconstructible charged particle tracks and around 100,000 to 120,000 3D space points (hits). The TrackML detector model consists of an inner pixel detector with four barrel layers (volume 8) and pixel endcaps (volumes 7 and 9), followed by a short-strip detector with four barrel layers (volume 13) and endcap disks (volumes 12, 14, and 16), and an outer long-strip detector with two barrel layers (volume 17) and endcap disks (volume 18). Thus, the central barrel region contains up to 10 effective cylindrical layers: four pixel, four short-strip, and two long-strip layers. Noise hits are added at a rate of about 5\% per layer, uniformly distributed across the detector surfaces, to simulate realistic imperfections such as inefficiencies or spurious signals\cite{Kiehn2019TrackML} . Ground truth information is provided for every hit, including the associated particle ID, momentum components, charge, production vertex coordinates, and whether the hit is reconstructible. The full dataset comprises 8850 training events split into 5 files and 125 test events, with a total size of approximately 100 GB for the training set alone.

\subsection{Detector Geometry and Particle Trajectories}
The detector geometry in the TrackML dataset approximates an all-silicon tracker similar to those in LHC experiments like CMS and ATLAS, with concentric barrel layers at increasing radii and forward/backward endcap disks to cover high pseudorapidity regions. Transverse momentum of a particle ($p_T$) is derived from the ground-truth particle information provided in the TrackML \texttt{particles.csv} file, calculated as $p_T = \sqrt{p_x^2 + p_y^2}$, where $p_x$ and $p_y$ are the momentum components in the transverse plane. $p_T$ information is mapped to the corresponding spacepoint hits and used to inform the physics-aware edge proposal and adaptive thresholding mechanisms. In this study, truth-level $p_T$ is used strictly as an idealized proxy to establish the framework for curvature-aware tracking. Ultimately, the architecture is designed to transition away from external momentum estimates, instead relying on internal geometric curvature mappings (such as the local ratio $\Delta\phi / \Delta r$) and framing the extraction of highly curved trajectories as a purely geometric problem the GNN must solve dynamically. \\
The radius of curvature $R$ of the helical path in the transverse plane is given by $R = p_T / (0.3 q B)$, where $p_T$ is the transverse momentum in GeV/$c$, $q$ is the particle charge in units of the elementary charge, and $B$ is the magnetic field strength in Tesla, yielding $R$ in meters \cite{pdg2024}. This formula arises from the balance between the centripetal force and the magnetic force, $p_T / R = q B$, with the constant 0.3 accounting for unit conversions (since $c = 3 \times 10^8$ m/s and 1 GeV/$c$ corresponds to appropriate scaling) \cite{pdg2024}. For the CMS Phase-2 upgrade, a similar geometry is planned with an Inner Tracker (pixel sensors) consisting of 4 barrel layers and 14 endcap disks per side, and an Outer Tracker (strip sensors) with 6 barrel layers and 15 endcap disks per side, covering a total active area of ~200 m$^2$ under a 3.8 T field \cite{apollo2020hl}. High-momentum particles ($p_T \gg 1$ GeV) leave relatively straighter tracks with minimal azimuthal deviations $\Delta\phi$ between consecutive layers, approximating linear propagation \cite{pdg2024}. In contrast, low-momentum particles ($p_T < 1$ GeV) exhibit pronounced curvature, resulting in large $\Delta\phi$ (up to several radians) and non-radial paths that deviate significantly from straight lines \cite{pdg2024}. Material interactions, such as multiple scattering (proportional to $1/p_T$) \cite{pdg2024} and energy loss via ionization (Bethe-Bloch formula) \cite{pdg2024}, further perturb trajectories, especially for low-$p_T$ particles. In high-pileup environments (140--200 interactions per crossing at HL-LHC) \cite{apollo2020hl}, overlapping tracks, secondary particles from decays or interactions, and noise complicate reconstruction. Causal constraints enforce valid connections only for increasing radius ($\Delta r > 0$, outward flow from the interaction point) and angular deviations consistent with the helical model, incorporating expected $\Delta\phi \approx (0.3 q B \Delta r)/p_T$ between layers separated by $\Delta r$ \cite{pdg2024}.

\subsection{Evaluation Metrics}
Performance is measured using similar scoring metrics used in TrackML studies\cite{Kiehn2019TrackML}, which balance reconstruction completeness and purity in high-density environments . A track is considered reconstructible if it originates from a charged particle has at least 5 associated hits, and satisfies detector acceptance criteria (e.g., originating near the primary vertex) . The tracking efficiency is defined as the fraction of such reconstructible truth tracks that are successfully matched to a reconstructed track, where matching requires that at least 50\% of the hits in the reconstructed track belong to the same truth particle (majority voting) . Conversely, the fake rate (or purity inverse) is the fraction of reconstructed tracks that are not matched to any single truth particle, i.e., where no truth particle contributes more than 50\% of the hits, often due to noise, mergers, or spurious connections \cite{Kiehn2019TrackML}. The overall score is computed as efficiency $\times$ (1 - fake rate), providing a single metric that rewards high efficiency while penalizing high fake rates, with a maximum value of 1 . To avoid double-counting, each truth track is matched to at most one reconstructed track, prioritizing the best match based on hit overlap. 

\subsection{Graph Formulation for Tracking}
In graph-based tracking approaches, hits are represented as nodes in a graph, with each node encoded by a feature vector capturing local and geometric information to facilitate learned pattern recognition \cite{DeZoort2023}. Typical node features include spatial coordinates in Cartesian ($x$, $y$, $z$) and cylindrical ($r$, $\phi$, $\eta$) systems, kinematic estimates when available , layer, volume, and module identifiers, particle charge, hit type (pixel or strip), and derived quantities like local angles or energy deposits \cite{DeZoort2023}. Edges represent potential track segments connecting hits, typically formed between hits in consecutive or nearby layers using heuristics such as k-nearest neighbors (kNN) in embedding spaces, radial expansion filters ($\Delta r > 0$), or physics-inspired constraints on $\Delta \phi$ and $\Delta \eta$ \cite{DeZoort2023}. In the Exa.TrkX pipeline and similar frameworks, initial graphs are built via metric learning, where a multi-layer perceptron (MLP) embeds nodes into a low-dimensional space, and edges connect nearest neighbors to sparsify the graph while preserving likely true connections \cite{DeZoort2023}. Message passing is performed using graph neural networks, such as interaction networks, where information propagates along edges: edge features (e.g., geometric differentials like $\Delta r$, $\Delta \phi$, $\Delta z$) are updated via relational MLPs, then aggregated at nodes to refine embeddings \cite{DeZoort2023}. Multiple layers of message passing enable global context awareness, capturing complex helical patterns and resolving ambiguities in dense regions \cite{DeZoort2023}. Final track reconstruction involves edge classification (binary true/fake scores) followed by graph segmentation, often using connected components on high-confidence edges or clustering algorithms like DBSCAN in the learned embedding space \cite{DeZoort2023}. For object condensation methods, nodes are clustered directly in latent space without explicit edge classification \cite{DeZoort2023}. This graph formulation leverages the sparse, irregular structure of detector data, enabling end-to-end differentiable learning of tracking patterns, with demonstrated linear scaling in computational complexity for large events \cite{DeZoort2023}.

\section{Methodology}
\label{sec:methodology}

\begin{figure*}[t]
    \centering
    \includegraphics[width=\textwidth]{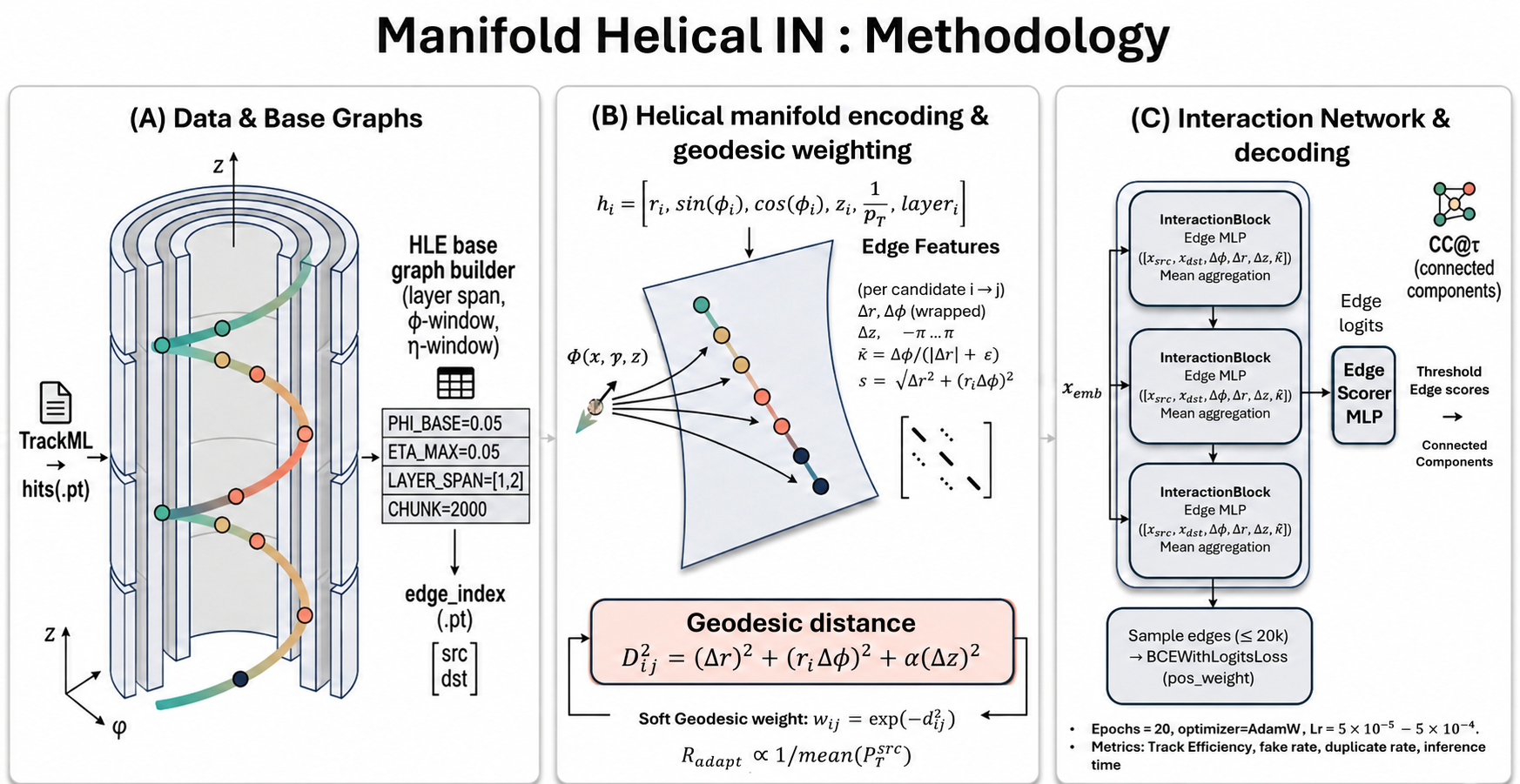}
    \caption{Architecture of the Manifold Helical-IN pipeline.
    (A) Candidate-graph construction from TrackML spacepoints using
    compatible detector layers and $p_T$-adaptive geometric constraints.
    (B) Helical node and edge encoding using periodic azimuthal features,
    local curvature information, and geometry-aware edge weighting.
    (C) Edge classification through stacked Interaction Blocks, followed
    by thresholding and connected-components track formation.}
    \label{fig:methodology}
\end{figure*}

The proposed \textit{Manifold Helical Interaction Network} (Manifold Helical-IN) forms track candidates from detector spacepoints by combining $p_T$-adaptive graph construction, curvature-aware geometric encoding, and weighted message passing. As illustrated in Fig.~\ref{fig:methodology}, the pipeline consists of three stages. First, a directed candidate graph is constructed using detector-layer compatibility and $p_T$-dependent geometric constraints. Second, the node and edge attributes are mapped to a helical representation containing periodic azimuthal coordinates, local curvature features, and geometry-aware edge weights. Third, stacked Interaction Blocks classify the candidate edges, after which thresholding and connected-components clustering produce the reconstructed track candidates.

\subsection{Data and Base-Graph Construction}
\label{subsec:graph_construction}

As illustrated in Fig.~\ref{fig:methodology}(A), each collision event is represented as a directed graph
\begin{equation}
    \mathcal{G}=(\mathcal{V},\mathcal{E}),
    \label{eq:tracking_graph}
\end{equation}
where $\mathcal{V}$ is the set of detector hits and
$\mathcal{E}\subseteq\mathcal{V}\times\mathcal{V}$ is the set of directed candidate edges. Each node $v_i\in\mathcal{V}$ corresponds to a detector hit with Cartesian coordinates $(x_i,y_i,z_i)$. The associated cylindrical coordinates are defined as
\begin{equation}
    r_i=\sqrt{x_i^2+y_i^2},
    \qquad
    \phi_i=\operatorname{atan2}(y_i,x_i),
    \qquad
    \eta_i=\operatorname{asinh}\left(\frac{z_i}{r_i}\right),
    \label{eq:cylindrical_coordinates}
\end{equation}
where $r_i$ is the radial distance from the beam axis, $\phi_i$ is the azimuthal angle, and $\eta_i$ is the spatial pseudorapidity. Each hit is also assigned a detector-layer index $L_i$ and the transverse momentum $p_{T,i}$ of its associated truth particle.

Prior to network training, an initial base graph is constructed statically for each event and stored in PyTorch Geometric format. Candidate edges are formed from a source hit $i$ in layer $L$ to a destination hit $j$ in layer $L' = L + \Delta L$, considering adjacent and next-to-adjacent layer transitions defined by the layer span:
\begin{equation}
    \Delta L_{ij} = L_j - L_i \in \{1, 2\}.
    \label{eq:layer_constraints}
\end{equation}
To account for the stronger magnetic curvature of low-$p_T$ trajectories, the angular candidate-search window in the azimuthal plane is dynamically scaled according to the transverse momentum of the source hit. The candidate set is restricted by a $p_T$-dependent azimuthal cut and a fixed pseudorapidity cut:
\begin{equation}
    \left|\Delta\phi_{ij}\right| \leq \phi_{\max}^{\text{dyn}}(p_{T,i}),
    \qquad
    \left|\Delta\eta_{ij}\right| \leq \eta_{\max},
    \label{eq:angular_constraints}
\end{equation}
where $\eta_{\max} = 0.05$, $\Delta\eta_{ij} = \eta_j - \eta_i$,
\begin{equation}
    \Delta\phi_{ij}
    =
    \operatorname{atan2}
    \left[
        \sin(\phi_j-\phi_i),
        \cos(\phi_j-\phi_i)
    \right]
    \label{eq:wrapped_phi}
\end{equation}
is the wrapped azimuthal difference in $[-\pi,\pi]$, and the dynamic azimuthal threshold is defined as:
\begin{equation}
    \phi_{\max}^{\text{dyn}}(p_{T,i}) = \Phi_{\text{base}} \cdot \left( \frac{C_{\text{scale}}}{\max(p_{T,i}, p_{\text{min}}) } \right),
    \label{eq:phi_max_dyn}
\end{equation}
with base window $\Phi_{\text{base}} = 0.05$, curvature scale factor $C_{\text{scale}} = 2.0$, and a minimum momentum clamping threshold $p_{\text{min}} = 0.1\text{ GeV/$c$}$. This inverse momentum scaling widens the azimuthal acceptance for lower-momentum particles experiencing larger bending radii.

Using this base-building pipeline across the TrackML dataset yields initial graphs containing an average of approximately $91{,}881$ nodes and $1.48 \times 10^6$ candidate edges per event. It is important to note that this statically generated structure is not the final graph used by the model; rather, it serves as an initial template, and the graph is further constructed and refined dynamically during the model's forward pass using manifold proximity and $p_T$-adaptive radius constraints, as detailed subsequently.

For supervised training, each candidate edge in this base structure is assigned a binary ground-truth label $y_{ij} \in \{0, 1\}$ based on particle association:
\begin{equation}
    y_{ij} = \begin{cases} 
    1 & \text{if } \text{pid}_i = \text{pid}_j \text{ and } \text{pid}_i \neq 0 \\ 
    0 & \text{otherwise} 
    \end{cases},
    \label{eq:edge_labeling}
\end{equation}
where $\text{pid}_i$ denotes the truth particle ID associated with hit $i$ (with background noise hits assigned $\text{pid} = 0$).

\subsection{Helical Manifold Encoding and Geometry Weighting}
\label{subsec:manifold_encoding}

As shown in Fig.~\ref{fig:methodology}(B), each detector hit is mapped to a curvature-aware representation that combines cylindrical geometry, periodic azimuthal coordinates, transverse momentum, pseudorapidity, and detector-layer information. The candidate edges are then described using local geometric differences and assigned learned geometry-aware weights.

\subsubsection{Dynamic Edge Evolution and Adaptive Search}
\label{subsec:dynamic_edges}

While the static base graph provides initial candidate connections across detector layers ($\Delta L \in \{1, 2\}$), the network dynamically proposes and refines edges during its forward propagation pass. To recover tightly curling low-$p_T$ trajectories that may deviate significantly, the candidate search window for adjacent layer transitions ($L \to L+1$) is adapted dynamically based on the source hit's transverse momentum. 

Using the latent space embeddings, pairwise Euclidean distances are computed between source and target nodes, and a $p_T$-adaptive search radius is enforced:
\begin{equation}
    R_{\mathrm{adapt}}(p_{T,i}) 
    = 
    \frac{3 r_{\mathrm{base}}}{p_{T,i}} + \frac{c}{p_{T,i}},
    \label{eq:adaptive_radius}
\end{equation}
where $r_{\mathrm{base}} = 0.08$ is the base radius, $c = 0.12$ is an empirical curvature penalty, and $p_{T,i}$ is clamped at a minimum of $0.1$ GeV/$c$. Candidate edges within this boundary that satisfy a helical angular compatibility mask are scored by an edge-scoring MLP, and only the top $K=10$ destination hits are retained per source node. These dynamically proposed edges are combined with the static base edges via unique union before message passing.\\
While Equation 8 currently parameterizes the adaptive search radius using transverse momentum, the underlying framework is fundamentally geometric. In a fully autonomous setting, this neighborhood boundary can be driven directly by the local curvature proxy $\hat{\kappa}_{ij}$ (Eq. 13) or local manifold proximity metrics, enabling the network to dynamically scale its search window for tightly curling trajectories without requiring explicit external momentum inputs.

\subsubsection{Node Encoding}

For hit $i$, we define the scaled quantities
\begin{equation}
    \widetilde{r}_i=\frac{r_i}{r_0},
    \qquad
    \widetilde{z}_i=\frac{z_i}{z_0},
    \qquad
    \widetilde{\eta}_i=\frac{\eta_i}{\eta_0},
    \qquad
    \widetilde{L}_i=\frac{L_i}{L_0},
    \label{eq:node_scaling}
\end{equation}
where $r_0=z_0=1000~\mathrm{mm}$, $\eta_0=5$, and $L_0=20$ are fixed scaling constants. The corresponding seven-dimensional node vector is
\begin{equation}
    \mathbf{u}_i=
    \left[
        \widetilde{r}_i,\,
        \sin\phi_i,\,
        \cos\phi_i,\,
        \widetilde{z}_i,\,
        \log\left(1+\frac{p_{T,i}}{p_0}\right),\,
        \widetilde{\eta}_i,\,
        \widetilde{L}_i
    \right],
    \label{eq:helical_node_features}
\end{equation}
where $p_0=1~\mathrm{GeV}$ makes the logarithm dimensionless. The pair
$\left(\sin\phi_i,\cos\phi_i\right)$ removes the coordinate discontinuity at
$\phi=\pm\pi$, while the logarithmic transformation compresses the dynamic range of $p_T$. The node vector is mapped to an initial latent representation through
\begin{equation}
    \mathbf{h}_i^{(0)}
    =
    f_{\mathrm{node}}(\mathbf{u}_i),
    \qquad
    \mathbf{h}_i^{(0)}\in\mathbb{R}^{d_h},
    \label{eq:node_embedding}
\end{equation}
where $f_{\mathrm{node}}$ is a multilayer perceptron with Gaussian error linear unit activations and $d_h$ is the latent dimension.

\subsubsection{Edge Encoding}

For every candidate edge $(i,j)\in\mathcal{E}$, we calculate the scaled radial and longitudinal differences
\begin{equation}
    \Delta\widetilde{r}_{ij}
    =
    \widetilde{r}_j-\widetilde{r}_i,
    \qquad
    \Delta\widetilde{z}_{ij}
    =
    \widetilde{z}_j-\widetilde{z}_i,
    \label{eq:scaled_edge_differences}
\end{equation}
together with the wrapped azimuthal difference
$\Delta\phi_{ij}$ defined in Eq.~\eqref{eq:wrapped_phi}. A local curvature proxy is defined as
\begin{equation}
    \widehat{\kappa}_{ij}
    =
    \frac{\Delta\phi_{ij}}
    {\left|\Delta\widetilde{r}_{ij}\right|+\epsilon_{\kappa}},
    \label{eq:local_curvature}
\end{equation}
where $\epsilon_{\kappa}=10^{-3}$ prevents division by zero. We also define the transverse path-length proxy
\begin{equation}
    s_{ij}
    =
    \sqrt{
        \left(\Delta\widetilde{r}_{ij}\right)^2
        +
        \left(\widetilde{r}_i\Delta\phi_{ij}\right)^2
    }.
    \label{eq:path_length}
\end{equation}
The geometric edge vector is then
\begin{equation}
    \mathbf{g}_{ij}
    =
    \left[
        \Delta\widetilde{r}_{ij},\,
        \Delta\phi_{ij},\,
        \Delta\widetilde{z}_{ij},\,
        \widehat{\kappa}_{ij},\,
        s_{ij}
    \right].
    \label{eq:geometric_edge_features}
\end{equation}

\subsubsection{Geometry-Aware Edge Weighting}

The geometric separation between connected hits is defined as
\begin{equation}
    d_{ij}^{2}
    =
    \left(\Delta\widetilde{r}_{ij}\right)^2
    +
    \left(\widetilde{r}_i\Delta\phi_{ij}\right)^2
    +
    \alpha
    \left(\Delta\widetilde{z}_{ij}\right)^2,
    \label{eq:geometric_distance}
\end{equation}
where $\alpha\geq0$ is a learned scalar controlling the contribution of longitudinal separation. This distance is converted into an edge weight according to
\begin{equation}
    w_{ij}=\exp\left(-d_{ij}^{2}\right),
    \qquad
    0<w_{ij}\leq1.
    \label{eq:geometry_weight}
\end{equation}
This formulation is referred to as a geodesic distance because it approximates the shortest path across the curved, unwrapped cylindrical manifold of the tracking detector layers, accounting for both radial depth and periodic azimuthal topology. The parameter $\alpha$ is enforced to be strictly non-negative during training using a Softplus activation function ($\alpha = \operatorname{Softplus}(\hat{\alpha})$) to prevent negative distance accumulation. Consequently, because $d_{ij}^2 \geq 0$ is strictly guaranteed, the exponential mapping in Eq.~\eqref{eq:geometry_weight} naturally bounds the edge weights such that $0 < w_{ij} \leq 1$.

Edges connecting geometrically aligned hits receive larger weights, while edges with greater geometric separation contribute less during message passing.

\subsection{Interaction Network and Track Decoding}
\label{subsec:interaction_network}

As shown in Fig.~\ref{fig:methodology}(C), the encoded node and edge features are processed by a sequence of three Interaction Blocks~\cite{Battaglia2016InteractionNetworks}. Each block updates the node representations by constructing edge messages, applying the geometry-aware weights, and aggregating the weighted messages at their destination nodes.

\subsubsection{Interaction Blocks}

Let $\mathbf{h}_i^{(\ell)}$ denote the latent representation of node $i$ at Interaction Block $\ell$, with $\mathbf{h}_i^{(0)}$ given by Eq.~\eqref{eq:node_embedding}. For each candidate edge $(i,j)\in\mathcal{E}$, an unweighted edge message is calculated as
\begin{equation}
    \widetilde{\mathbf{m}}_{ij}^{(\ell)}
    =
    f_{\mathrm{edge}}^{(\ell)}
    \left(
        \mathbf{h}_i^{(\ell)}
        \mathbin{\|} 
        \mathbf{h}_j^{(\ell)}
        \mathbin{\|}
        \mathbf{g}_{ij}
    \right),
    \label{eq:edge_message}
\end{equation}
where $f_{\mathrm{edge}}^{(\ell)}$ is the edge multilayer perceptron in block $\ell$, $\mathbf{g}_{ij}$ is the geometric edge vector defined in Eq.~\eqref{eq:geometric_edge_features}, and $\mathbin{\|}$ denotes vector concatenation. The message is modulated by the geometry-aware weight:
\begin{equation}
    \mathbf{m}_{ij}^{(\ell)}
    =
    w_{ij}\widetilde{\mathbf{m}}_{ij}^{(\ell)},
    \label{eq:weighted_message}
\end{equation}
where $w_{ij}$ is defined in Eq.~\eqref{eq:geometry_weight}.

The incoming messages at destination node $j$ are aggregated using
\begin{equation}
    \overline{\mathbf{m}}_j^{(\ell)}
    =
    \frac{1}{|\mathcal{N}(j)|}
    \sum_{i\in\mathcal{N}(j)}
    \mathbf{m}_{ij}^{(\ell)},
    \label{eq:message_aggregation}
\end{equation}
where
\begin{equation}
    \mathcal{N}(j)
    =
    \left\{i\in\mathcal{V}:(i,j)\in\mathcal{E}\right\}
    \label{eq:incoming_neighbors}
\end{equation}
is the set of source nodes connected to node $j$. The destination-node representation is then updated as
\begin{equation}
    \mathbf{h}_j^{(\ell+1)}
    =
    f_{\mathrm{node}}^{(\ell)}
    \left(
        \mathbf{h}_j^{(\ell)}
        \mathbin{\|}
        \overline{\mathbf{m}}_j^{(\ell)}
    \right),
    \label{eq:node_update}
\end{equation}
where $f_{\mathrm{node}}^{(\ell)}$ is the node-update multilayer perceptron. Repeating this procedure across three blocks allows each node representation to incorporate information from increasingly distant hits in the candidate graph.

\subsubsection{Edge Scoring}

After the final Interaction Block, each candidate edge is assigned a logit
\begin{equation}
    a_{ij}
    =
    f_{\mathrm{score}}
    \left(
        \mathbf{h}_i^{(L)}
        \mathbin{\|}
        \mathbf{h}_j^{(L)}
        \mathbin{\|}
        \mathbf{g}_{ij}
    \right),
    \label{eq:edge_logit}
\end{equation}
where $L=3$ is the number of Interaction Blocks and $f_{\mathrm{score}}$ is the edge-scoring multilayer perceptron. The predicted edge probability is
\begin{equation}
    \widehat{y}_{ij}
    =
    \sigma(a_{ij})
    =
    \frac{1}{1+\exp(-a_{ij})},
    \label{eq:edge_probability}
\end{equation}
where $\sigma(\cdot)$ is the sigmoid function and
$\widehat{y}_{ij}\in(0,1)$ estimates the probability that edge $(i,j)$ belongs to a true particle trajectory.

\subsubsection{Connected-Components Track Formation}

Candidate edges whose predicted probability exceeds a threshold $\tau_{ij}$ are retained:
\begin{equation}
    \mathcal{E}_{\tau}
    =
    \left\{
        (i,j)\in\mathcal{E}:
        \widehat{y}_{ij}\geq\tau_{ij}
    \right\}.
    \label{eq:thresholded_edges}
\end{equation}
The threshold can depend on the transverse-momentum range, as specified in Sec.~\ref{sec:experimental_setup}. For track formation, the retained edges are treated as undirected connections, and connected-components clustering partitions the thresholded graph into hit sets
\begin{equation}
    \left\{\mathcal{C}_1,\ldots,\mathcal{C}_M\right\}
    =
    \operatorname{CC}
    \left(\mathcal{V},\mathcal{E}_{\tau}\right),
    \label{eq:connected_components}
\end{equation}
where $\mathcal{C}_k\subseteq\mathcal{V}$ is the set of hits assigned to reconstructed track candidate $k$, and $M$ is the number of reconstructed candidates.

\section{Experimental Setup}
\label{sec:experimental_setup}

\subsection{Dataset and Preprocessing}

The experiments utilize a total of 3,894 labeled events drawn from the TrackML dataset. To build the candidate graphs, 3,540 events were sequentially sampled from the sub-releases of the dataset. These were randomly partitioned using a fixed seed into 3,186 training events (90\%) and 354 validation events (10\%) for model selection and threshold tuning. Final evaluation and benchmark reporting were conducted on a strictly unseen set of 354 events sampled independently from the TrackML test dataset, ensuring no cross-contamination of detector hits. Furthermore, unless otherwise specified, all tables and performance metrics reported in Section V are derived solely from this unseen test set. The events are divided at the event level into disjoint training, validation, and test sets to prevent hits from the same collision event from appearing in multiple subsets. Following preprocessing, each event contains an average of approximately 91\,881 nodes and 1.48 million candidate edges.

Candidate graphs connect hits from adjacent and next-to-adjacent detector layers, corresponding to $\Delta L\in\{1,2\}$. Edge proposals are restricted using
\begin{equation}
    |\Delta\phi_{ij}|\leq 0.05,
    \qquad
    |\Delta\eta_{ij}|\leq 0.05,
\end{equation}
and at most $K=10$ destination hits are retained for each source hit. Truth-level $p_T$ is mapped from each simulated particle to its associated hits and used by the $p_T$-adaptive graph-construction procedure.

\subsection{Model and Training Configuration}

The Helical-IN baseline and Manifold Helical-IN use three Interaction Blocks. Both models use a node-embedding dimension of 64, a message-passing hidden dimension of 128, and an edge-embedding dimension of 32. The Manifold Helical-IN uses the seven-dimensional node representation defined in Eq.~\eqref{eq:helical_node_features}, while the baseline uses the standard spatial representation without manifold encoding or geometry-aware edge weighting.

To ensure a rigorous and fair comparison, the hyperparameters for both the baseline Helical-IN and the proposed Manifold Helical-IN were optimized independently using grid search on the validation set. As a result, the models employ different optimizers (Adam vs. AdamW), learning rates, and positive-class weights (8.0 vs. 5.0) as summarized in Table~\ref{tab:hyperparams}. This independent tuning guarantees that the baseline is evaluated at its peak capability for the dataset, isolating the geometric manifold embedding as the true driver of the observed performance gains.

\begin{table}[htpb]
    \centering
    \caption{Model and training configuration.}
    \label{tab:hyperparams}
    \setlength{\tabcolsep}{5pt}
    \begin{tabular}{lcc}
        \toprule
        \textbf{Parameter}
        & \textbf{Helical-IN}
        & \textbf{Manifold Helical-IN} \\
        \midrule
        Optimizer & Adam & AdamW \\
        Learning rate & $5\times10^{-5}$ & $5\times10^{-4}$ \\
        Weight decay & $1\times10^{-5}$ & $1\times10^{-5}$ \\
        Epochs & 20 & 20 \\
        Batch size & 1 & 1 \\
        Positive class weight & 8.0 & 5.0 \\
        Interaction Blocks & 3 & 3 \\
        Node-embedding dimension & 64 & 64 \\
        Message-passing hidden dimension & 128 & 128 \\
        Edge-embedding dimension & 32 & 32 \\
        Gradient-clipping norm & 1.0 & 1.0 \\
        \bottomrule
    \end{tabular}
\end{table}

All experiments are performed on a single NVIDIA A100 GPU with 80~GB of memory using PyTorch Geometric and automatic mixed-precision computation.

\subsection{Evaluation Protocol}

A reconstructed track candidate is matched to the truth particle contributing more than 50\% of its hits. If several truth particles contribute hits, the particle with the largest hit overlap is selected. Reconstruction performance is measured using efficiency, fake-track rate, mean hit recovery fraction (MHRF), and inference latency.

Let $\mathcal{T}$ denote the set of selected truth particles and $\mathcal{R}$ the set of reconstructed track candidates. The reconstruction efficiency is
\begin{equation}
    \mathrm{Efficiency}
    =
    \frac{
        \left|
        \left\{
        t\in\mathcal{T}:
        \exists r\in\mathcal{R}
        \text{ matched to }t
        \right\}
        \right|
    }{|\mathcal{T}|}.
    \label{eq:efficiency}
\end{equation}
The fake-track rate is
\begin{equation}
    \mathrm{Fake\ Rate}
    =
    \frac{
        \left|
        \left\{
        r\in\mathcal{R}:
        r\text{ is unmatched}
        \right\}
        \right|
    }{|\mathcal{R}|}.
    \label{eq:fake_rate}
\end{equation}

A connected component is classified as a valid reconstructed track candidate if it contains a minimum of $N_{\mathrm{hits}} \geq 5$ spacepoints; isolated hits and two-hit fragments are discarded as noise prior to evaluation. A reconstructed track is successfully matched to a truth particle if its purity $P > 0.50$, meaning more than 50\% of the track's hits originate from that single simulated particle (majority voting). 

In cases of combinatorial ambiguity:
\begin{itemize}
    \item \textbf{Duplicates:} If multiple reconstructed tracks match the same truth particle, the candidate with the highest hit recovery is classified as the true positive, while the others are penalized as duplicates.
    \item \textbf{Fakes (Merged Tracks):} Reconstructed tracks containing hits from multiple particles where no single particle achieves $P > 0.50$ are counted strictly as fake tracks.
\end{itemize}

For each matched truth particle $t$, let $H(t)$ denote its truth-hit set and let $C_{\mathrm{best}}(t)$ denote the reconstructed candidate containing the largest number of hits from $H(t)$. The MHRF is
\begin{equation}
    \mathrm{MHRF}
    =
    \frac{1}{|\mathcal{T}_{\mathrm{matched}}|}
    \sum_{t\in\mathcal{T}_{\mathrm{matched}}}
    \frac{
        |H(t)\cap C_{\mathrm{best}}(t)|
    }{|H(t)|},
    \label{eq:mhrf}
\end{equation}
where $\mathcal{T}_{\mathrm{matched}}\subseteq\mathcal{T}$ is the set of matched truth particles.

For a scalar summary of efficiency and fake-track suppression, we also calculate
\begin{equation}
    S
    =
    \mathrm{Efficiency}
    \left(1-\mathrm{Fake\ Rate}\right).
    \label{eq:combined_score}
\end{equation}

Inference latency is measured per event on the A100 GPU and includes candidate-graph construction, GNN inference, edge thresholding, and connected-components clustering. GPU synchronization is applied before and after each timed execution, and the reported latency is averaged across the evaluation events after warm-up runs.

\section{Results}
\label{sec:results}

\subsection{Reconstruction Performance}
\label{subsec:reconstruction_performance}

Manifold Helical-IN substantially improves track purity while maintaining high reconstruction efficiency for particles with $p_T>0.5~\mathrm{GeV}$. The proposed model reduces the fake-track rate from $0.0508$ to $0.0037$, corresponding to a reduction of $92.7\%$ or a factor of $13.7$. Its reconstruction efficiency reaches $0.9675$, compared with $0.9717$ for the Helical-IN baseline, a difference of $0.0042$. The results summarized in Table~\ref{tab:reconstruction_performance} show that the manifold representation strongly suppresses false track candidates with only a small change in reconstruction efficiency.

\begin{table}[htbp]
    \centering
    \caption{Reconstruction performance for particles with $p_T > 0.5$ GeV on test set}
    \label{tab:reconstruction_performance}
    \begin{tabular}{lcc}
        \toprule
        \textbf{Model}
        & \textbf{Efficiency}
        & \textbf{Fake-track rate} \\
        \midrule
        Helical-IN
        & \textbf{0.9717}
        & 0.0508 \\
        Manifold Helical-IN
        & 0.9675
        & \textbf{0.0037} \\
        \bottomrule
    \end{tabular}
\end{table}

\subsection{Momentum-Dependent Performance}
\label{subsec:momentum_performance}

\begin{figure}[t]
    \centering
    \includegraphics[width=0.85\linewidth]{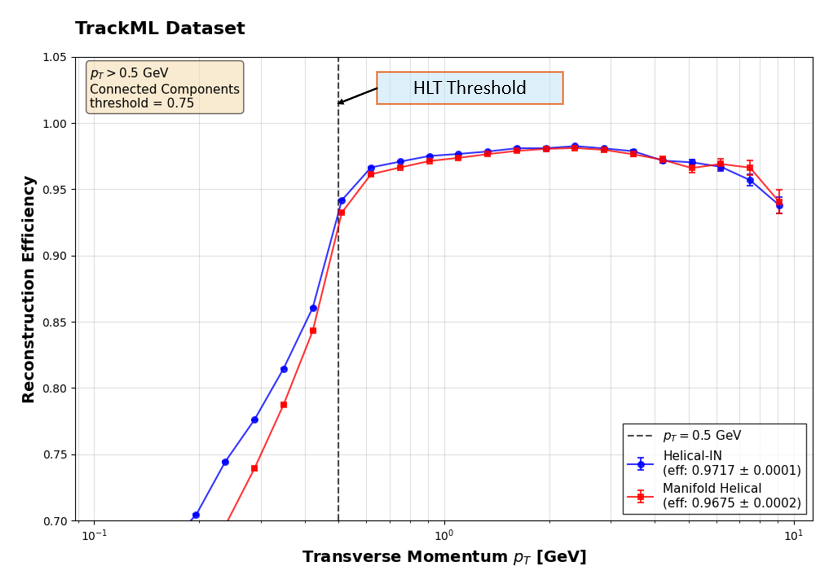}
    \caption{Reconstruction efficiency as a function of particle
    transverse momentum $p_T$ for Helical-IN and Manifold Helical-IN.
    Both models reach a high-efficiency plateau above approximately
    $1~\mathrm{GeV}$, while reconstruction becomes more difficult at
    lower transverse momentum.}
    \label{fig:ptvseff}
\end{figure}

The reconstruction efficiency of both models increases strongly with transverse momentum, as shown in Fig.~\ref{fig:ptvseff}. In the low-momentum interval $0.1<p_T<0.5~\mathrm{GeV}$, the Helical-IN baseline achieves an average efficiency of $0.783$, compared with $0.746$ for Manifold Helical-IN. The difference decreases as $p_T$ increases, and both models enter a high-efficiency regime above approximately $0.5~\mathrm{GeV}$. For $p_T>1~\mathrm{GeV}$, the baseline and proposed model achieve efficiencies of $0.978$ and $0.976$, respectively.

These results show that Manifold Helical-IN preserves the strong high-$p_T$ reconstruction performance of the baseline. The remaining efficiency loss at low $p_T$ is mainly associated with the initial candidate graph, where strongly curved trajectories may fail the layer and angular edge-proposal constraints. The main gain from the manifold representation is therefore its suppression of false track candidates, as shown in Sec.~\ref{subsec:reconstruction_performance}.

\subsection{Ablation Study}
\label{sec:ablation_study}

To isolate the contribution of each architectural component, we conduct a systematic ablation study. We evaluate four distinct model variants: the baseline Helical-IN, the fully proposed Manifold Helical-IN, and two ablated versions of our architecture. Table \ref{tab:ablation_study} summarizes the global performance metrics across the validation set. The results reveal a clear division of labor between the core innovations. The Manifold Embedding serves as the primary engine for track purity, drastically suppressing combinatorial noise. Conversely, the $p_T$-adaptive search radius acts as a specialized mechanism to recover tightly curling, low-momentum trajectories.

\begin{table}[htbp]
\centering
\caption{Ablation Study: Component-wise Analysis on the TrackML Validation Set.}
\label{tab:ablation_study}
\renewcommand{\arraystretch}{1.2}
\begin{tabular}{lccc}
\toprule
\textbf{Model Variant} & \textbf{Eff. ($p_T>0.5$)}& \textbf{Fake Rate} & \textbf{Time (ms)} \\
\midrule
Helical-IN (Baseline) & 0.967& 0.0508 & 150.4 \\
\textbf{Full Manifold Helical-IN} & \textbf{0.968}& \textbf{0.0037} & \textbf{150.6} \\
Ablation 1 (Standard Emb.) & 0.977& 0.0480 & 155.7 \\
Ablation 2 (Fixed Radius) & 0.968& 0.0052 & 150.4 \\
\bottomrule
\end{tabular}
\end{table}

The most striking result emerges from Ablation 1. This variant replaces the helical manifold projection with a standard 11-dimensional Cartesian node embedding while retaining the physics-based edge masks. The 11 baseline features consist of raw coordinates ($x, y, z$), cylindrical transformations ($r, \phi, \eta$), truth momentum ($p_T$), detector metadata ($\text{layer\_id}, \text{module\_id}$), and hit properties ($q, n_{\mathrm{hits}}$). 

Counterintuitively, this baseline achieves the highest overall reconstruction efficiency of 0.977. However, without the strict geometric constraints of the manifold chart, the network loses its ability to distinguish true helices from spurious intersections, causing the fake rate to explode from 0.0037 to 0.0480. Furthermore, Ablation 2 investigates the necessity of the curvature-aware search window by replacing the dynamic radius with a fixed, dimensionless upper bound ($R_{\mathrm{adapt}} = 0.8$). While maintaining a low fake rate (0.0052), the fixed window degrades reconstruction efficiency in the critical $0.1 < p_T < 0.5$ GeV turn-on region, as shown in Table \ref{tab:ablation_study}.

\begin{figure}[htbp]
\centering
\includegraphics[width=\textwidth]{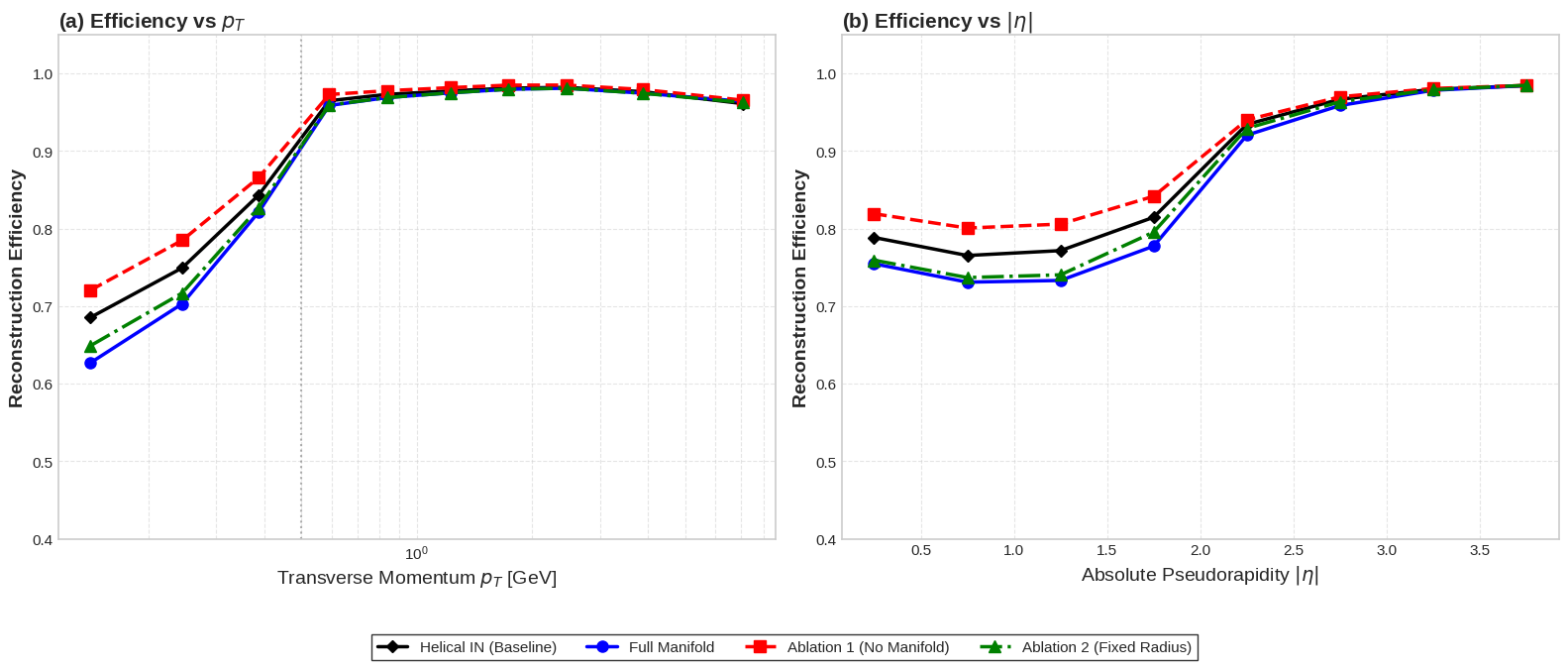}
\caption{Kinematic dependence of the ablation variants. \textbf{(a)} Reconstruction efficiency as a function of transverse momentum ($p_T$), illustrating the impact of the adaptive radius on the low-$p_T$ turn-on curve. \textbf{(b)} Efficiency versus absolute pseudorapidity ($|\eta|$), demonstrating the geometric robustness of the full model across the central barrel and forward endcap regions.}
\label{fig:ablation_kinematic}
\end{figure}

Furthermore, the pseudorapidity dependence in Figure \ref{fig:ablation_kinematic}(b) highlights that the adaptive radius ensures uniform performance across the entire detector geometry. It seamlessly bridges the dense central barrel and the forward endcaps. Finally, all four variants process events in approximately 150 ms. This confirms that integrating explicit geometric priors introduces zero computational overhead, preserving the high-throughput requirements of modern collider experiments.

\subsection{Evaluation Protocol}
\label{sec:eval_protocol}

Inference is executed using a highly optimized, vectorized pipeline. After the model computes edge probabilities, a curvature-adaptive thresholding mechanism is applied: edges connecting hits with an average $p_T < 0.5$ GeV are retained if their score exceeds 0.45, while higher-momentum edges require a score of at least 0.65. The surviving edges are clustered into track candidates using sparse graph connected components algorithm , which processes complex events in approximately 150 ms.

Performance is quantified using the metrics defined in Section 2.3. While overall efficiency is calculated across the full $p_T > 0.1$ GeV spectrum, Our primary focus of this study is to evaluate performance on lower $P_T$ tracks, So a threshold of $p_T > 0.5$ GeV is chosen to align with the kinematic boundaries critical for soft Quantum Chromodynamics (QCD) physics and heavy-flavor decay reconstruction, providing a fair and experimentally relevant comparison with contemporary Transformer and Point Transformer architectures.

\begin{figure}[h]
\centering
\includegraphics[width=0.8\textwidth]{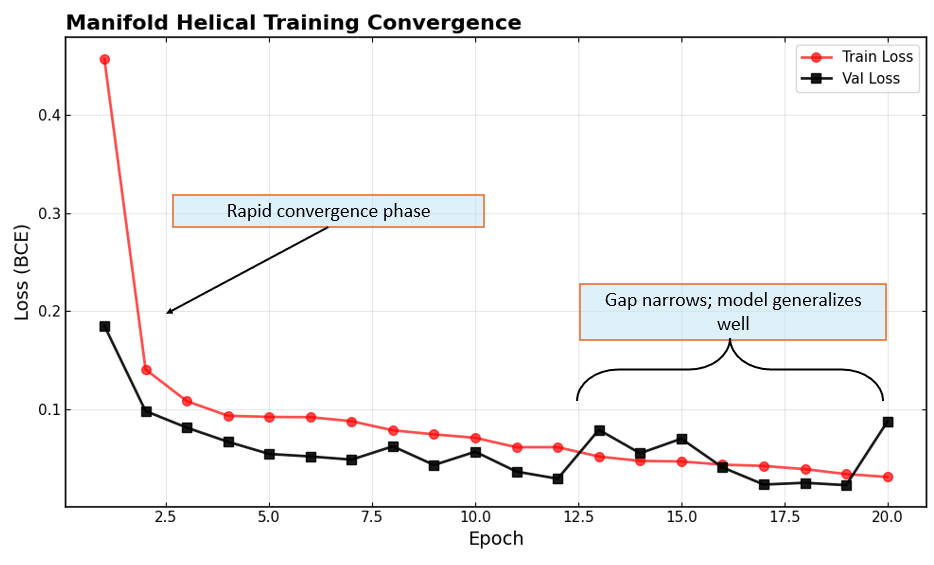}
\caption{Training dynamics of the Manifold Helical-IN. Convergence of Binary Cross-Entropy loss on the training and validation sets over 20 epochs, demonstrating stable learning without overfitting.}
\label{fig:training_dynamics}
\end{figure}

\subsection{Global Performance Benchmarks}
\label{sec:global_benchmarks}

Table \ref{tab:global_benchmarks} summarizes the global performance benchmarks for both architectures across the full TrackML validation set. When evaluated over the entire kinematic spectrum ($p_T > 0.1$ GeV), the baseline Helical-IN achieves an overall efficiency of 0.8592, while the Manifold Helical-IN yields 0.8357. This overall metric is inherently constrained by the notoriously difficult low-momentum regime ($0.1 < p_T < 0.5$ GeV), where severe multiple scattering and tight curling induce combinatorial ambiguities. 

However, in the range ($p_T > 0.5$ GeV),  both architectures achieve an impressive reconstruction efficiency of 0.9675. Crucially, the Manifold Helical-IN distinguishes itself through good track purity, suppressing the fake rate to a mere 0.0037—a drastic reduction from the baseline's 0.0508. Furthermore, the proposed model maintains good trajectory coherence, evidenced by a high Mean Hit Recovery Fraction (MHRF) of 0.858, and processes complex events in approximately 150 ms, matching the computational throughput of the baseline without requiring specialized hardware acceleration.
\begin{table}[h]
\centering
\caption{Global Performance Benchmarks on TrackML (Truth-Filtered). The Manifold Helical-IN achieves a massive reduction in Fake Rate while maintaining comparable track coherence (MHRF) and inference latency.}
\label{tab:global_benchmarks}
\begin{tabular}{lcccc}
\toprule
\textbf{Model} & \textbf{Efficiency} & \textbf{Fake Rate} & \textbf{MHRF} & \textbf{Latency (ms)} \\
\midrule
Helical-IN (Baseline) & 0.8592 & 0.0508 & 0.8745 & 150.36 \\
\textbf{Manifold Helical-IN (Ours)} & \textbf{0.8357} & \textbf{0.0037} & \textbf{0.8579} & \textbf{150.63} \\
\bottomrule
\end{tabular}
\end{table}

\subsection{Momentum-Dependent Reconstruction and Low-$p_T$ Performance}
Figure \ref{fig:ptvseff} illustrates the reconstruction efficiency as a function of transverse momentum ($p_T$). Both architectures exhibit good performance in the standard tracking regime ($p_T > 1.0$ GeV), with the Helical-IN achieving 0.978 efficiency and the Manifold Helical-IN closely matching it at 0.976. 

Figure~4 shows the momentum dependence of the reconstruction efficiency for both architectures. To provide a more quantitative view of the low-$p_T$ turn-on and the transition into the high-efficiency regime, Table~\ref{tab:pt_efficiency} reports the efficiency in finer $p_T$ bins from 0.1 to 10~GeV. The degradation below approximately 0.5~GeV is pronounced for both models, reflecting the increasing curvature and combinatorial ambiguity of low-momentum trajectories. In contrast, the efficiency rises rapidly above 0.5~GeV and reaches a stable plateau above approximately 1~GeV, where the Manifold Helical-IN achieves efficiencies above 97\%.

The bin populations are also important for interpreting the inclusive efficiency. A substantial fraction of the reconstructed particle population lies below 1~GeV, so the lower efficiency in this region has a significant effect on the full-spectrum result. This motivates reporting both the inclusive performance over the full evaluated momentum range and the $p_T>0.5$~GeV benchmark used for comparison with recent high-quality tracking studies. The detailed binning also makes clear that the reduced inclusive efficiency is not representative of the model's performance in the higher-momentum tracking regime.

In the challenging low-$p_T$ regime ($0.1 - 0.5$ GeV), where tight curling and multiple scattering degrade performance, the Helical-IN baseline maintains a slight efficiency advantage (0.783 vs. 0.746). However, the primary advantage of the Manifold Helical-IN is not raw low-$p_T$ efficiency, but rather its good purity. As shown in Table 3, the Manifold embedding reduces the Fake Rate by an order of magnitude (from 0.0508 to 0.0037). This results in a higher overall TrackML score ($\text{Eff} \times (1 - \text{Fake})$) of \textbf{0.8326} for the Manifold model, compared to 0.8156 for the baseline, demonstrating that the manifold projection successfully suppresses combinatorial noise without sacrificing high-$p_T$ tracking capability.

\subsection{Track Coherence and the MHRF Metric}
To rigorously evaluate track fragmentation, we utilize the Mean Hit Recovery Fraction (MHRF), which measures the percentage of a particle's true hits contained within its primary reconstructed cluster. 

As demonstrated in Table 3, the proposed Manifold Helical-IN maintains a high MHRF of \textbf{0.858}, nearly identical to the Helical-IN baseline (0.874). This confirms that the integration of the helical manifold embedding with the strict Interaction Network message-passing successfully preserves global trajectory continuity. This also indicates that the manifold embedding substantially preserves hit continuity while suppressing spurious combinatorial connections. 

\begin{figure}[h]
    \centering
    % Placeholder: Insert the Multiplicity vs Eta plot
    \includegraphics[width=0.9\linewidth]{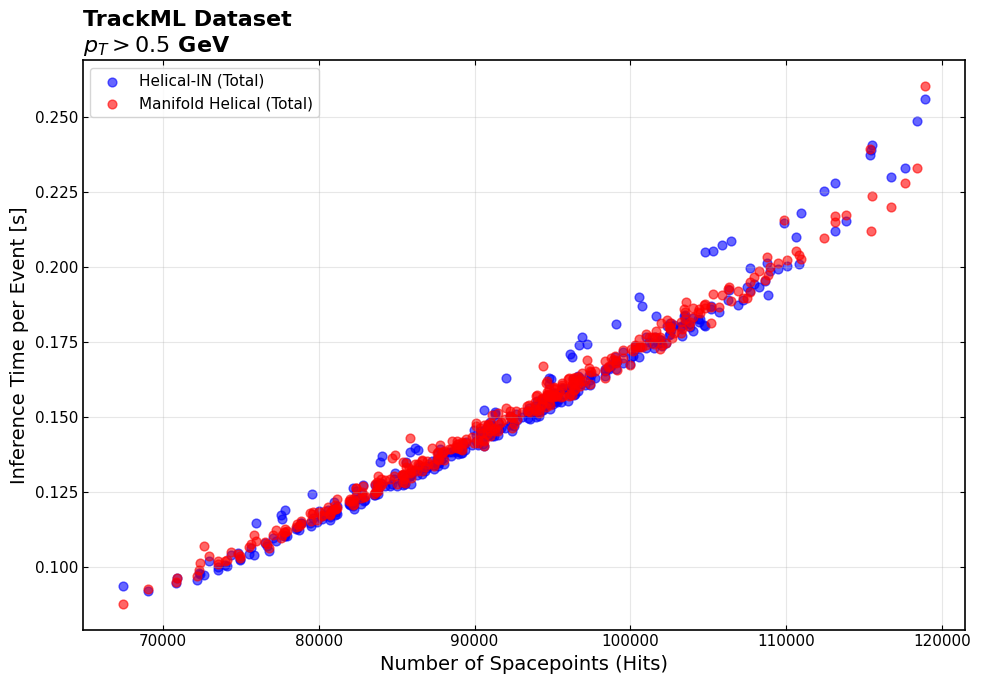}
    \caption{Inference time per event.}
    \label{fig:multiplicity}
\end{figure}

\subsection{Comparison with Recent Approaches}
\label{sec:sota_comparison}

To contextualize the performance of the proposed architectures, we compare our models with several recent tracking approaches, including the foundational Exa.TrkX graph neural network~\cite{Ju2021ExaTrkX}, the 2025 Transformer Tracker~\cite{VanStroud2025Transformer}, the EggNet pipeline~\cite{Chan2025EggNet}, and the recent HEPTv2 Point Transformer~\cite{Miao2026HEPTv2}.

A key distinction lies in the kinematic regime under consideration. Many contemporary methods apply relatively high transverse-momentum thresholds (typically $p_T > 0.9$--$1.0$\,GeV) and report results primarily in that easier domain. Even if not while training, almost no study shows explicitly the result against $P_T$ tracks of all regime. In contrast, we evaluate performance down to $p_T > 0.1$\,GeV with steadily improving efficiency from  $p_T > 0.3$\,GeV, a region that remains important for soft QCD processes and heavy-flavor decay reconstruction and that presents a substantially higher combinatorial challenge due to stronger track curvature.

In the range $p_T > 0.5$\,GeV, both the baseline Helical-IN and the Manifold Helical-IN achieve a reconstruction efficiency of 0.9675. This is comparable to the efficiencies reported by recent Transformer-based trackers (approximately 0.94--0.97) and approaches the 0.986 efficiency of the highly optimized HEPTv2 architecture, while covering a more demanding low-$p_T$ range than is commonly emphasized in those works.

Track purity shows a clearer differentiation. The Manifold Helical-IN reaches a fake rate of 0.0037 for $p_T > 0.5$\,GeV. This is substantially lower than the fake rates typically reported for Exa.TrkX (approximately 0.03--0.05), the 2025 Transformer Tracker (0.007--0.012), and EggNet (0.0019–0.0021). The baseline Helical-IN records a higher fake rate of 0.0508. These results indicate that the combination of a helical manifold embedding with $p_T$-aware geometric filtering is particularly effective at suppressing combinatorial noise in the low-to-moderate $p_T$ region.

Computational cost is also relevant for high-luminosity applications. Both of our models process full TrackML events in approximately 150\,ms on a single NVIDIA A100 GPU using standard dense operations and sparse connected-component clustering. For reference, the full EggNet pipeline requires roughly 1.47\,s per event on the same hardware, largely due to iterative clustering, while HEPTv2 achieves substantially lower latency (approximately 15\,ms) through specialized sparse-attention kernels. It should be noted that the published timing figures for EggNet and HEPTv2 correspond to evaluations performed with higher transverse-momentum thresholds ($p_T > 1$\,GeV and $p_T > 900$\,MeV, respectively). In contrast, the 150\,ms latency reported here is measured on a sample that includes the more combinatorially challenging low-$p_T$ region. Our approach therefore offers a practical trade-off between purity, low-$p_T$ coverage, and moderate computational requirements without relying on highly customized kernels.

\begin{table*}[htbp]
\centering
\caption{Focused comparison in the experimentally relevant high-quality tracking regime ($p_T > 0.5$ GeV).}
\label{tab:sota_comparison}
\renewcommand{\arraystretch}{1.15}
\small
\begin{tabular}{p{2.8cm} p{2.2cm} c c c c p{2.6cm}}
\toprule
\textbf{Method} & \textbf{$p_T$ Range} & \textbf{Eff.} & \textbf{Fake} & \textbf{Latency} & \textbf{Hardware} \\
\midrule
Exa.TrkX \cite{Ju2021ExaTrkX} 
& $> 0.5$ GeV 
& $88.7$ 
& $0.03$--$0.05$ 
& $2.2 \pm0.3$ s 
& V100 (typ.) \\

Transformer Tracker \cite{VanStroud2025Transformer} 
& $>750$ MeV 
& $0.94$--$0.97$ 
& $0.007$--$0.012$ 
& $97$ ms 
& V100 / A100 \\

EggNet \cite{Chan2025EggNet} 
& $>1$ GeV 
& $\sim 0.96$ 
& $0.0019$--$0.0021$ 
& $1.47$ s 
& A100 GPU \\

HEPTv2 \cite{Miao2026HEPTv2} 
& > 0.9 GeV 
& $\sim 0.986$ 
& $\sim 0.008$ 
& $15$ ms 
& A100 GPU \\

\midrule
\textbf{Helical-IN (Ours)} 
& $>0.5$ GeV 
& \textbf{0.9717} 
& \textbf{0.0508} 
& \textbf{150 ms} 
& A100 GPU \\

\textbf{Manifold (Ours)} 
& $>0.5$ GeV 
& \textbf{0.9675} 
& \textbf{0.0037} 
& \textbf{150 ms} 
& A100 GPU \\
\bottomrule
\end{tabular}
\end{table*}
\section{Discussion}
\label{sec:discussion}

The integration of a helical manifold embedding fundamentally alters the geometric representation used for track reconstruction, providing a robust mechanism to suppress combinatorial noise in dense environments. As demonstrated by our benchmarks, the manifold projection achieves a dramatic reduction in the fake-track rate (from 0.0508 to 0.0037) while preserving high reconstruction efficiency (0.9675) for particles with $p_T > 0.5$ GeV. 

A central focus of our ongoing development is aligning this computational throughput with real-time High-Level Trigger (HLT) constraints. Processing full events in approximately 150 ms. This confirms that our approach is optimized for heterogeneous, GPU-accelerated HLT computing farms rather than custom FPGA-based Level-1 trigger boards, directly addressing the computational demands of capturing soft, unclustered signatures like SUEPs.

To further scale these capabilities for the High-Luminosity LHC, our next phase of research focuses on two primary architectural evolutions:
\begin{enumerate}
    \item \textbf{Autonomous Dynamic Curvature Estimation:} We are actively replacing the idealized truth-$p_T$ proxy for the dynamic edge evolution with self-contained, continuous local curvature metrics ($\hat{\kappa}_{ij}$) computed directly during message passing. This will allow the adaptive search radius ($R_{adapt}$) to evolve purely from local manifold geometry, entirely eliminating external momentum inputs.
    \item \textbf{Hybrid Transformer-GNN Architectures:} Recognizing that high-$p_T$ trajectories ($p_T > 2$ GeV) in the outer and middle detector layers are relatively straight and well-suited for fast Transformer attention mechanisms, we are developing a hybrid pipeline. In this paradigm, Transformers handle outer-tracker reconstruction, while the Manifold Helical-IN is deployed exclusively in the inner pixel layers to untangle the tightly curling, highly combinatorial 500 MeV trajectories.
\end{enumerate}

\section{Conclusion}
\label{sec:conclusion}

In this work, we introduced the Manifold Helical-IN, a physics-informed graph neural network designed to tackle the combinatorial complexities of low-momentum track reconstruction. By embedding detector spacepoints onto a helical manifold chart, the model leverages curvature-aware features and geodesic-weighted message passing to suppress spurious background tracks by an order of magnitude, achieving a fake rate of 0.0037 in the $p_T > 0.5$ GeV regime. Operating at an inference latency of 150 ms per event on an NVIDIA A100 GPU, the architecture provides a viable pathway toward real-time soft track reconstruction at the HLT. While rigid layer-to-layer constraints currently limit ultra-low $p_T$ efficiency, our ongoing work on arc-length adjacency, local dynamic curvature, and hybrid Transformer-GNN integration aims to resolve these bottlenecks. These developments establish a robust, modular framework for recovering fragile Beyond Standard Model signatures in the high-occupancy environment of the HL-LHC.

\bibliography{biblio}

\begin{table}[htbp]
\centering
\caption{Reconstruction efficiency in fine-grained $p_T$ bins. Reported uncertainties represent $\pm 1\sigma$ binomial statistical standard errors based on the event count per kinematic bin.}
\label{tab:pt_efficiency}
\begin{tabular}{lccc}
\hline
$p_T$ Range (GeV) & Helical-IN Eff & Manifold Eff & Particles/Bin\\
\hline
$0.10 - 0.12$ & $0.0730 \pm 0.0075$ & $0.0187 \pm 0.0056$ & 1{,}068 \\
$0.12 - 0.15$ & $0.0885 \pm 0.0074$ & $0.0338 \pm 0.0068$ & 1{,}480 \\
$0.15 - 0.18$ & $0.6810 \pm 0.0008$ & $0.6270 \pm 0.0012$ & 332{,}797 \\
$0.18 - 0.22$ & $0.7043 \pm 0.0006$ & $0.6428 \pm 0.0010$ & 494{,}708 \\
$0.22 - 0.26$ & $0.7444 \pm 0.0006$ & $0.6963 \pm 0.0009$ & 573{,}597 \\
$0.26 - 0.32$ & $0.7760 \pm 0.0005$ & $0.7392 \pm 0.0008$ & 631{,}618 \\
$0.32 - 0.38$ & $0.8147 \pm 0.0005$ & $0.7875 \pm 0.0007$ & 666{,}384 \\
$0.38 - 0.46$ & $0.8606 \pm 0.0004$ & $0.8431 \pm 0.0006$ & 658{,}874 \\
$0.46 - 0.56$ & $0.9418 \pm 0.0003$ & $0.9325 \pm 0.0005$ & 579{,}406 \\
$0.56 - 0.68$ & $0.9665 \pm 0.0003$ & $0.9614 \pm 0.0004$ & 510{,}029 \\
$0.68 - 0.83$ & $0.9709 \pm 0.0003$ & $0.9665 \pm 0.0004$ & 427{,}618 \\
$0.83 - 1.00$ & $0.9751 \pm 0.0003$ & $0.9713 \pm 0.0004$ & 332{,}865 \\
$1.00 - 1.21$ & $0.9767 \pm 0.0003$ & $0.9737 \pm 0.0004$ & 252{,}836 \\
$1.21 - 1.47$ & $0.9785 \pm 0.0003$ & $0.9765 \pm 0.0005$ & 187{,}968 \\
$1.47 - 1.78$ & $0.9810 \pm 0.0004$ & $0.9790 \pm 0.0005$ & 136{,}225 \\
$1.78 - 2.15$ & $0.9810 \pm 0.0004$ & $0.9805 \pm 0.0006$ & 92{,}115 \\
$2.15 - 2.61$ & $0.9826 \pm 0.0005$ & $0.9811 \pm 0.0008$ & 57{,}846 \\
$2.61 - 3.16$ & $0.9808 \pm 0.0007$ & $0.9798 \pm 0.0010$ & 34{,}463 \\
$3.16 - 3.83$ & $0.9787 \pm 0.0011$ & $0.9764 \pm 0.0016$ & 19{,}039 \\
$3.83 - 4.64$ & $0.9717 \pm 0.0016$ & $0.9723 \pm 0.0023$ & 10{,}481 \\
$4.64 - 5.62$ & $0.9704 \pm 0.0022$ & $0.9660 \pm 0.0034$ & 5{,}836 \\
$5.62 - 6.81$ & $0.9670 \pm 0.0029$ & $0.9691 \pm 0.0041$ & 3{,}394 \\
$6.81 - 8.25$ & $0.9569 \pm 0.0044$ & $0.9663 \pm 0.0055$ & 2{,}276 \\
$8.25 - 10.00$ & $0.9380 \pm 0.0063$ & $0.9406 \pm 0.0088$ & 1{,}598 \\
\hline
\end{tabular}
\end{table}
\end{document}